\documentstyle[12pt]{article}
\newcommand{\be}{\begin{equation}}
\newcommand{\ee}{\end{equation}}
\newcommand{\ba}{\begin{eqnarray}}
\newcommand{\ea}{\end{eqnarray}}

\begin{document}
\begin{center}
{\bf\Huge  {  Quantization  of Floreanini-Jackiw chiral harmonic
oscillator  }}
\end{center} \begin{center} Dumitru Baleanu\footnote{ Permanent address
: Institute of Space Sciences, P.O.BOX, MG-36, R
76900,Magurele-Bucharest, Romania,E-Mail
address:~~baleanu@thsun1.jinr.ru,baleanu@venus.ifa.ro}\\ Bogoliubov
Laboratory of Theoretical Physics \\ Joint Institute for Nuclear
Research\\ Dubna, Moscow Region, Russia

\end{center}
\begin{center}
    and
\end{center}
\begin{center}
 Yurdahan $G\ddot{u}ler$ \footnote{E-Mail address:~~yurdahan@ari.cankaya.edu.tr}
\end{center}
\begin{center}
Department of Mathematics and Computer Sciences, Faculty of Arts and Sciences,
Cankaya University,Ankara,Turkey
\end{center}

\vskip 5mm
\bigskip
\nopagebreak
\begin{abstract}
The Floreanini-Jackiw formulation of the chiral quantum-mechanical system
oscillator  is a model of constrained theory with only second-class constraints.
in the Dirac's classification.The covariant quantization needs infinite number
of auxiliary variables and a Wess-Zumino term.
In this paper we investigate the path integral quatization of this model
using $G\ddot{u}ler's$ canonical formalism.
All variables  are  gauge variables in  this formalism.
The  Siegel's action is obtained using Hamilton-Jacobi formulation of
the systems with constraints.
\end{abstract}

\section{Introduction }
Chiral bosons in two-dimensional space-time and $(2+1)$ -dimensional
Chern-Simons(CS) gauge theories are related problems which have been
attracting much attention.These problems are important for the string
 program and for the development of the quantum Hall
 effect\cite{gross},\cite{wen}.Floreanini  and Jackiw suggested an
 action suitable for the quantization of a two-dimensional chiral
boson\cite{floreanini}.Siegel proposed an apparently unrelated action
for the same system\cite{siegel}.
 In\cite{jack} the connection between these two approaches was
 investigated. The quantization of Siegel action was investigated by
Faddeev-Jackiw formalism in\cite{faddeev}.

The equivalent Lagrangian method is used to obtain the
set of Hamilton-Jacobi partial differential equations \cite{gu87}.In other
words the equations of motion are written as total differential
equations in many variables.

The main aim of this paper is to investigate the quantization of the
Floreanini-Jackiw chiral oscillator using $G\ddot{u}ler's$ formalism.

The plan of our paper is the following:

In Section 2 we present the quantization  of the field theories with
constraints using Hamilton-Jacobi method.
In Section 3 the path integral quantization for Floreanini-Jackiw chiral
oscillator is given.The Siegel's action was obtained using
$G\ddot{u}ler's$ formalism.
In Section 4 we present our conclusions.

\section{Hamilton-Jacobi quantization of the field theories with
constraints}
Starting from Hamilton-Jacobi partial-differential
equation the singular systems was investigated using a formalism
introduced by
$G\ddot{u}ler$(see for example Refs.\cite{gu87},\cite{gubal}).

 The canonical formulation  gives the set of Hamilton-Jacobi
partial-differential equation as
\be
H_{\alpha}^{'}(\chi_{\beta},\phi_{\alpha},{\partial S\over\partial
\phi_{\alpha}},{\partial S\over\partial \chi_{\alpha}})=0,
\alpha,\beta=0,n-r+1,\cdots,n,a=1,\cdots,n-r,
\ee
where
\be\label{(ham)}
H_{\alpha}^{'}=H_{\alpha}(\chi_{\beta},\phi_{a},\pi_{a}) +\pi_{\alpha}
\ee
and $H_{0}$ is the canonical hamiltonian.
 The equations of motion are obtained as total differential equations
in many variables as follows
\be\label{(pq)}
d\phi_{a}={\partial H_{\alpha}^{'}\over\partial \pi_{a}}d\chi_{\alpha},
d\pi_{a}=-{\partial H_{\alpha}^{'}\over\partial
q_{\alpha}}d\chi_{\alpha}, d\pi_{\mu}=-{\partial
H_{\alpha}^{'}\over\partial \chi_{\mu}}d\chi_{\alpha}, \mu=1,\cdots, r
\ee \be\label{(z)} dz=(-H_{\alpha} +\pi_{a}{\partial
H_{\alpha}^{'}\over\partial \pi_{a}})d\chi_{\alpha} \ee where
$z=S(\chi_{\alpha},\phi_{a})$.The set of
equations(\ref{(pq)},\ref{(z)}) is integrable if \be\label{(h1)}
dH_{0}^{'}=0,dH_{\mu}^{'}=0,\mu=1,\cdots r
\ee
If conditions(\ref{(h1)}) are not satisfied identically , one considers
them as a new constraints and again tests the consistency conditions.
Thus repeating this procedure one may obtain a set of conditions.

Let suppose that  for a system with constraints we
found all independent hamiltonians $H_{\mu}^{'}$  using the
calculus of variations \cite{gu87},\cite{gubal}.At this
stage we will use Dirac's procedure of
quatization \cite{Dirac}.  We have
\be
H_{\alpha}^{'}\Psi=0,\mu=1,\cdots,r
\ee
where $\Psi$ is the wave
function.
The consistency conditions are
\be\label{(cond)}
[H_{\alpha}^{'},H_{\beta}^{'}]\Psi=0,\alpha,\beta=1,\cdots ,r
\ee
If  the hamiltonians $H_{\alpha}^{'}$
satisfies
\be
[H_{\alpha}^{'},H_{\beta}^{'}]=C_{\alpha\beta}^{\gamma}H_{\gamma}^{'}
\ee
they are of first class in the Dirac's classification.
On the other hand if
\be
[H_{\alpha}^{'},H_{\beta}^{'}]=C_{\alpha\beta}
\ee
where $C_{\alpha\beta}$ do not depend of $\phi_{i}$ and $\pi_{i}$
 then from(\ref{(cond)})
there arises naturally  Dirac's brackets and the canonical
quatization will be performed taking Dirac's brackets into
commutators.

$G\ddot{u}ler's$ formalism gives an action when
all hamiltonians $H_{\alpha}^{'}$  are in involution.Since in
this  formalism we work from the beginning in the extended
space we suppose that variables $t_{\alpha}$ depend of $\tau$.Here
$\tau$ is canonical conjugate with $p_{0}$.

If we are able , for a given
 system with constraints, to find the independent hamiltonians
$H_{\alpha}^{'}$ in involution then we can perform the quantization of
this system using path integral quantization method with the action
given by (\ref{(z)})

\be\label{(z2)}
z=\int(-H_{\alpha} +\pi_{\beta}{\partial H_{\alpha}^{'}\over\partial
\pi_{\beta}})\dot{\chi_{a}}d{\tau}
\ee
where
$\dot\chi_{\alpha}={d\chi_{\alpha}\over d\tau}$.

\section{Chiral Oscillator }
We consider the  Lagrangian\cite{flor}
\be\label{(lag)}
L_{0}=\omega{\dot{q_{i}}^{(0)}}\epsilon_{ij}q_{j}^{(0)}+\omega^{2}q_{i}^{(0)}
q_{i}^{(0)},i,j=1,2
\ee
{ }From(\ref{(lag)}) we found the constraints
\be
\Omega_{i}=p_{i}^{(0)} - \omega\epsilon_{ij}q_{j}^{(0)},i=1,2
\ee

and the canonical hamiltonian
\be
H_{c}=(p_{i}^{(0)}-\omega\epsilon_{ij}q_{j}^{(0)}){\dot q_{i}}^{(0)} -
{\omega}^{2}q_{k}^{(0)}q_{k}^{(0)}
\ee
Then in the $G\ddot{u}ler's$ formalism we have  following
 hamiltonians \be H_{0}^{'}=p_{0} +H_{c},H_{i}^{'}=p_{i}^{(0)}
-\omega\epsilon_{ij}q_{j}^{(0)},i=1,2
\ee
and all the variables $q_{i}^{(0)}$ are gauge variables.
The hamiltonians are not in involution because
\be
[H_{i}^{'},H_{j}^{'}]=-2\omega\epsilon_{ij}
\ee
In order to obtain the hamiltonians in involution
we will extend the space with new variable $p_{i}^{(1)}$ and
$q_{i}^{(1)}$.
The new expressions for the hamiltonians $H_{i}^{''}$ in involution are
\be
H_{i}^{''}=H_{i}^{'} -\omega\epsilon_{ij}q_{j}^{(1)} -p_{i}^{(1)}
\ee
but we get a new set of constraints
\be
H_{i}^{1'}=p_{i}^{(1)} -\omega\epsilon_{ij}q_{j}^{(1)}
\ee
If we repeat the procedure after N steps we get N+1 hamiltonians in
involution and  the hamiltonians $H_{i}^{N'}=p_{i}^{(N)}
-\omega\epsilon_{ij}q_{j}^{(N)}$ fulfilling
 \be
[H_{i}^{N'},H_{j}^{N'}]=-2\omega\epsilon_{ij}
\ee

The final form of the canonical hamiltonian obtained after an
infinite repetation of the conversion process is
\be
H_{c}^{(\infty)}=\Sigma_{k=0}^{\infty}(p_{i}^{(k)}-\omega\epsilon_{ij}{q_{i}^{(k)}}){\dot q_{i}^{k}}-\Sigma_{k=0}^{\infty}\omega^{2}q_{i}^{(k)}q_{i}^{(k)}
-
2\Sigma_{k=1}^{\infty}\Sigma_{m=0}^{k-1}\omega\epsilon_{ij}q_{j}^{(m)}
({\dot q_{i}^{(m)}} +\omega\epsilon_{il}q_{l}^{(m)})
\ee
Then in the $G\ddot{u}ler's$ formalism we have  infinite numbers of
hamiltonians in involution
\be
H_{0}^{'}=p_{0} +H_{c}^{(\infty)},H_{k}^{'}=p_{i}^{(k)}-
\omega\epsilon_{ij}q_{j}^{(k)},k=1,\cdots,\infty
\ee
Using (\ref{(z2)}) we found after some  calculations that the action has
the form
\be
z=\int{L d\tau}
\ee
where L is given by
\be
L=\Sigma_{k=0}^{\infty}(\omega\epsilon_{ij}{q_{i}^{(k)}}{\dot
q_{j}^{k}}+ \omega^{2}q_{i}^{(k)}q_{i}^{(k)}) +
2\Sigma_{k=1}^{\infty}\Sigma_{m=0}^{k-1}\omega\epsilon_{ij}q_{j}^{(m)}
({\dot q_{i}^{(m)}} +\omega\epsilon_{il}q_{l}^{(m)})
\ee
This result is in agreement with those from \cite{flor}.

\subsection{Siegel's action}
Siegel was first
one who proposed an action for the  chiral-boson problem \cite{siegel}.
The Lagrangean density is
\be
L={\partial_{-}\phi}{\partial_{+}\phi} +\lambda({\partial_{-}}\phi)^{2}
\ee
and the canonical hamiltonian becomes
\be
H_{c}={1\over 2}(1+\lambda)^{-1}(\pi +\lambda{\phi}^{'})^{2} +
{1\over 2}(1-\lambda)(\phi^{'})^{2}
\ee
where $\pi$ is the canonical momentum conjugate to $\phi$.
On the other hand we observe that
\be
\pi_{\lambda}=0
\ee
In $G\ddot{u}ler's$ formalism we have the following hamiltonians
\be\label{(ham1)}
H_{0}^{'}=p_{0} +H_{c}, H_{1}^{'}=\pi_{\lambda}
\ee
Imposing $dH_{0}^{'}=0$ and $dH_{1}^{'}=0$ we generate another
hamiltonian $H_{2}^{'}$ as
\be\label{(h2)}
H_{2}^{'}=\pi_{\phi}-{\phi}^{'}
\ee
{ }From (\ref{(ham1)}) and (\ref{(h2)}) we conclude that $\lambda$ and
$\phi$ are gauge variables in this formalism.

Now we are interested in performing the  quantization of the system.
Using Dirac's procedure we have
\be
H_{0}^{'}\Psi=0 ,H_{1}^{'}\Psi=0,H_{2}^{'}\Psi=0
\ee
Because
\be
[H_{1}^{'},H_{0}^{'}]={1\over 2}(1+\lambda)^{-1}(\pi-\phi^{'})^{2}
\ee
and
\be
[H_{2}^{'},H_{2}^{'}]={\partial_{x}}\delta(x-y)
\ee
we conclude that the system has second class constraints in the Dirac's
classification and we can quantize it using path integral quantization
method \cite{senj}.
Because we have obtained  the same constraints $H_{0}^{'}$,$H_{1}^{'}$
and $H_{2}^{'}$ as in \cite{jack}  we found  the same  result after
performing path integral quantization.

\section{Concluding remarks}
Using Hamilton-Jacobi formalism for systems with constraints we found
that Floreanini-Jackiw  chiral harmonic oscillator  is a theory with an
infinite number of Hamiltonians in involution.
The path integral quantization, using the action given by
$G\ddot{u}ler's$ formalism ,was performed and the results are in
agreement with those obtained by others authors .

For  Floreanini-Jackiw chiral harmonic oscillator
 we found that all the fields  are gauge fields in the $G\ddot{u}ler's$
formalism.
Because all  hamiltonians  $H_{\alpha}^{'}$ are
constraints in the extended space in
this  formalism we have no
first and second class constraints in the  Dirac's classification
 at the classical level.

However the first and second class
constraints become important in this formalism in the process of
quantization .
For the Siegel's action we found only three independent
hamiltonians $H_{0}^{'}$ ,$H_{1}^{'}$, $H_{2}^{'}$.This set of
hamiltonians give us the correct result if the path integral
quantization for this system is performed.

In  the
$G\ddot{u}ler's$ formalism  all the variations of constraints
$H_{0}^{'}$, $H_{1}^{'}$ and $H_{2}^{'}$ do not give us new
constraints.

The problem if the constraint $(H_{2}^{'})^{2}$(for
more details see Ref.\cite{jack})  is of first or second class
does not arise in the $G\ddot{u}ler's$ formalism.
   In this case we found that all variables are gauge variables.

 \section{Acknowledgements}

One of the authors (D.B.) would like to thank TUBITAK for financial
support  and METU for the hospitality during his
working stage at Department of Physics.


\begin{thebibliography}{99}
\bibitem{gross}D.J.Gross, J.A.Harvey, E.Martinec and R.Rohm,\sl
{Phys.Rev.Lett.}{\bf 54}, (1985) 502.
\bibitem{wen}X.G.Wen,{\sl Phys.Rev.Lett.}{\bf
64},2206(1990);\sl{Phys.Rev.B}{\bf 41},(1990) 1238.
\bibitem{floreanini}R.Floreanini and R.Jackiw,{\sl Phys.Rev.Lett.}{\bf
59},(1987) 1873.
\bibitem{siegel}W.Siegel,\sl{Nucl.Phys.}{\bf B 238},(1984) 307.
\bibitem{jack}M.Bernstein and J.Sonnenschein,\sl{Phys.Rev.Lett.}
{\bf 60},(1988) 1772.
\bibitem{faddeev}
L.Faddeev and R.Jackiw ,\sl{Phys.Rev.Lett.},{\bf 60}, (1988) 1692.

\bibitem{gu87}Y.$G\ddot{u}ler$,\sl{Il Nuovo Cimento B}, {\bf 100}
(1987) 251,
\sl{Il Nuovo Cimento B},{\bf 100} , (1987), 267,
\sl{J.Math.Phys.}{\bf 30} (1989), 785.
\bibitem{gubal}D.Baleanu and Y.$G\ddot{u}ler$, Hamiltonian-Jacobi
of the fields with constraints,sent to Il Nuovo Cimento B.
\bibitem{Dirac}Dirac P.A.M, \sl{Lectures on Quantum Mechanics }
(Yeshiva University, New York, N.Y.) 1964. \bibitem{flor}N.M.Horta
Barreira and C.Wotzasek,\sl{Phy.Rev.D.}{\bf 45,no.4} (1992) 1410.
 \bibitem{senj}Senjanovic P.,\sl{Ann.Phys.(N.Y.)},{\bf 100} (1976) 227.
\end{thebibliography}
\end{document}